\documentclass{article}
\usepackage{spconf,amsmath,graphicx,hyperref}
\usepackage{float}
\usepackage{subcaption} 
\usepackage{comment}
\usepackage{booktabs}

\title{Low-Latency Assistive Audio Enhancement for Neurodivergent People}
\twoauthors
 {Alexander Popescu\sthanks{The first author performed the work during his internship at Logitech}}
	{EPFL, Switzerland}
 {Rosie Frost, Milos Cernak}
	{Logitech Europe, Switzerland}
\begin{document}
\ninept
\maketitle
\begin{abstract}
Neurodivergent people frequently experience decreased sound tolerance, with estimates suggesting it affects 50–70\% of this population. This heightened sensitivity can provoke reactions ranging from mild discomfort to severe distress, highlighting the critical need for assistive audio enhancement technologies
In this paper, we propose several assistive audio enhancement algorithms designed to selectively filter distressing sounds.
To address this, we curated a list of potential trigger sounds by analyzing neurodivergent-focused communities on platforms such as Reddit. Using this list, a dataset of trigger sound samples was compiled from publicly available sources, including FSD50K and ESC50. These samples were then used to train and evaluate various Digital Signal Processing (DSP) and Machine Learning (ML) audio enhancement algorithms. Among the approaches explored, Dynamic Range Compression (DRC) proved the most effective, successfully attenuating trigger sounds and reducing auditory distress for neurodivergent listeners.
\end{abstract}

\section{Introduction}
From daily activities like attending school or going to work to more recreational pursuits such as watching movies or playing video games, people are routinely exposed to various sounds. In some individuals -- particularly those who are neurodivergent -- these sounds may cause distress, anxiety, or other negative reactions that can diminish overall quality of life. Indeed, Decreased Sound Tolerance (DST) is a well-documented condition among neurodivergent populations, affecting an estimated 50–70\% of individuals, although these figures remain subject to debate~\cite{williams_review_2021,yilmaz_assessment_2017,ren_prevalence_2021}. Neurodivergent people tend to be more sensory and can become easily triggered and dysregulated by sound, sight, smell, and touch. They have higher discomfort with their audio devices than the average person.

DST is the condition where everyday sounds that do not usually bother/annoy most people become abnormally bothersome in some way. It encompasses three major categories, namely Misophonia, Hyperacusis, and Phonophobia. These categories are believed to exhibit relatively high comorbidity among neurodivergent populations, particularly individuals with autism and Attention-Deficit/Hyperactivity Disorder (ADHD)~\cite{williams_review_2021,williams_prevalence_2021,abramovitch_neuropsychological_2024,rinaldi_autistic_2023}. Misophonia is typically characterized by strong negative emotional responses -- such as severe disgust, distress, or anxiety -- to specific ``trigger'' sounds. These triggers frequently include orofacial sounds (e.g., chewing, swallowing) but may also involve other commonplace, often repetitive, or transient noises encountered in daily life~\cite{andermane_phenomenological_2023,hansen_what_2021}. Hyperacusis, in contrast, involves heightened pain or sensitivity to specific frequencies, mostly high-frequency sounds~\cite{aazh_prevalence_2018,sheldrake_audiometric_2015}, while Phonophobia is characterized by anxiety in response to certain loud noises. The latter can be acquired following Hyperacusis or Misophonia, particularly after repeated intense exposures~\cite{williams_review_2021}. 
The reactions induced by these conditions can be quite severe and impair overall quality of life, underscoring the importance of developing assistive audio enhancement technologies for neurodivergent people.

Today's most assistive audio devices are noise-cancelling headphones/earbuds and hearing protection devices. Research has shown that noise cancellation can help neurodivergent individuals manage their decreased sound tolerance~\cite{pfeiffer_effectiveness_2019,neave_knowledge_2021}, improving focus and promoting a sense of calm. However, this avoidance strategy (all sounds suppression, including speech) comes with certain drawbacks, particularly for individuals with autism. It may limit them in essential activities and social interactions, potentially affecting communication and social skills~\cite{neave_knowledge_2021}. Additionally, while noise cancellation can provide relief, excessive reliance on it may hinder the natural development of coping mechanisms~\cite{pfeiffer_effectiveness_2019}.


A potential solution is having Active Noise Cancellation (ANC) that selectively generates tailored anti-sound, filtering out only trigger sounds while preserving non-triggering ones.
In a standard transparency mode, external sounds are simply played back by the headphones. By contrast, a selective transparency mode would combine active noise cancellation (ANC), which ideally filters out all external noise, with the output of an audio enhancement algorithm specifically designed to suppress triggering features of the outside sound.
However, this may not be feasible due to the strict latency constraints of ANC. Recent low-latency audio Machine Learning (ML) can achieve a few milliseconds~\cite{wang2022stft, wu2024ultra} and sub-milliseconds~\cite{dementyev2024towards} latency, but generating ANC anti-sound must occur here within a few nanoseconds, making ML-based algorithms too slow.

A viable alternative is to use ANC and a selective transparency mode that selectively plays back only non-triggering sounds from the outside. While ANC suppresses all outside audio, an audio enhancement method could process external audio, filtering out or attenuating trigger sounds before playing them back in the headphones alongside ANC, similar to the semantic hearing setup ~\cite{veluri_semantic_2023}, which focuses on extracting specific sounds from real-world environments and also playing them back alongside ANC.

This paper proposes several assistive audio enhancement algorithms that could be used on top of ANC to achieve a selective transparency mode. In addition to selective transparency mode, the proposed algorithms could also be used as audio plugins or apps on computers and phones, enhancing the device's audio before it reaches the speaker. Such enhancement applications could be beneficial for neurodivergent individuals during activities like gaming or watching movies.

The paper is structured as follows. Section~\ref{sec:methods} describes the methods and the experiments, including novel audio neurodivergent data construction, and Section~\ref{sec:results} presents and discusses the obtained results. Finally, Sec.~\ref{sec:conclusion} concludes the paper and outlines the future work.



\section{Methods and experimental setup}
\label{sec:methods}
The methodology consists of three major steps: i) design and construction of novel neurodivergent-trigger and neutral sound collections, ii) mixing trigger sounds and non-triggering neutral sounds for training and assessment of different audio processing algorithms, and iii) training and optimizing existing (baseline) algorithms on their ability to attenuate or completely filter out trigger sounds, using both objective metrics as well as subjective listening tests.

\subsection{Dataset creation}

In the first step, a list of trigger sounds was created, as no publicly available dataset contained a large or diverse enough collection of trigger sounds relevant to neurodivergent individuals. This list was constructed by extracting information from online neurodivergent communities on Reddit\footnote{\url{https://www.reddit.com/}}, particularly through a targeted scraping of neurodivergent-related subreddits (like *autism*, ADHD*, neurodiversity, etc.) using the PRAW API~\footnote{\url{https://github.com/praw-dev/praw}} that generated a set of about 11600 posts relevant to sound intolerance,
combined with format enforced, quantized Llama-3.1-8B-Instruct model from HuggingFace
to systematically extract and organize information about commentators' potential sound sensitivities.
This list was then mapped to the AudioSet ontology leaf-node labels using GPT-4, which enabled us to extract relevant audio samples from datasets such as FSD50K~\cite{fonseca_fsd50k_2022}, DISCO~\cite{lanzendorfer_disco-10m_2023} and ESC50~\cite{piczak_esc_2015} to create a collection of trigger audio samples and non-triggering/neutral audio samples, using the 25 most frequently mentioned labels. This resulted in approximately 5 hours of trigger sound audio. Note that the non-triggering sounds were collected only with labels that did not share the same subtree on the AudioSet ontology as any trigger labels.

Given collected triggering and neutral sounds, two datasets of sound mixtures of 10 seconds with distinct mixing mechanics were created: \textbf{Dataset 1} files consist of one trigger and one neutral sound mixed at an SNR of 0 dB, and one background ambient sound mixed at a lower SNR of -10 dB, whereas \textbf{Dataset 2} employed a randomized SNR between trigger and neutral sounds, with neutral sounds having an SNR ranging from -15dB to 5dB. For both datasets, the trigger was repeated for the full length of the mixture. Furthermore, both datasets were divided into training, validation, and testing, and it was ensured that no audio samples used in the training mixtures appeared in the validation or testing sets. Both datasets contained 20'000 mixtures for training, 1'000 for validation, and 1'000 for testing, whereas the test contained unseen trigger and non-trigger files as well as unseen base backgrounds.

\subsection{Assistive audio enhancement algorithms}

\subsubsection{DSP algorithms}

We identified and refined five DSP algorithms due to their straightforward implementation and favourable latency performance: i) Dynamic Range Compression (DRC), ii) Equalization (EQ), iii) Automatic Gain Control (AGC), iv) Multichannel Transient Noise Reduction (MCTR), and v) Low pass filter (LPF). We adopted literature-based parameters for the last two algorithms, MCTR and LPF. The parameters of DRC, EQ, and AGC were optimized on the validation set of \textbf{Dataset 2} to maximize SI-SNR. We employed Optuna’s framework~\cite{noauthor_optuna_nodate} to optimize the parameters through the Tree-structured Parzen Estimator (TPE) 
~\cite{watanabe_tree-structured_2023}.

We designed assistive audio enhancement as selective transparency mode that runs on top of ANC. 
DRC can be very useful in reducing transient loud sounds, which are known to be problematic for neurodivergent individuals, and it is usually causal, making it suitable for real-time applications. To implement DRC, we used the 
Spotify's Pedalboard library\footnote{\url{https://github.com/spotify/pedalboard}}
and optimized parameters, including the threshold, ratio, attack time, and release time using the same dataset as described above. The optimization process yielded optimal parameters of a threshold of -35 dB, a ratio of 30:1, an attack time of 0.01 ms, and a release time of 100 ms.
EQ adjusts the gain of different frequency bands~\cite{noauthor_equalization_2025}, which could be useful to attenuate certain (high) frequency features that might be triggering for neurodivergent individuals~\cite{aazh_prevalence_2018,sheldrake_audiometric_2015}. The equalizer was developed by using a combination of shelving filters~\cite{valimaki_all_2016} with the Pedalboard library.
The resulting optimal configuration included a gain of -8 dB at 200 Hz (low-shelf), -2.75 dB Hz (high-shelf), +1.6 dB at 5000 Hz (high-shelf), -3 dB at 10000 Hz (high-shelf), and -6 dB at 15000 Hz (high-shelf).
AGC automatically adjusts the volume, and it could help to lower the volume in the occurrence of sudden loud sounds, which might be useful for neurodivergent people who can be more sensitive to sudden loud noises. It is already widely used in hearing devices to regulate the flux of sudden loud noise and allow more comfort for a hearing aid user~\cite{keshavarzi_transient_2021}. The parameters that were fine-tuned on the data were the attack and release coefficients, as well as the target level and maximum gain. However AGC’s fine-tuning resulted in poor SI-SNR performance; we speculated it was caused by the target power level parameter.
In this work, we also applied the MCTR algorithm, developed by Keshvarzi et al.~\cite{keshavarzi_evaluation_2018}, which is a real-time audio processing method designed to reduce the loudness of transient sounds.
This multi-channel approach supposedly ensures that the unwanted transient peaks that are notoriously problematic for neurodivergent individuals are reduced while maintaining the natural quality and audibility of the overall audio, all with low latency suitable for real-time applications. We used the same parameters as the ones proposed by~\cite{keshavarzi_evaluation_2018}.
Neurodivergent people suffering from hyperacusis often perceive certain frequency ranges as significantly louder. In particular, frequencies between 1 kHz and 8 kHz appear much more pronounced compared to those without hyperacusis \cite{aazh_prevalence_2018,sheldrake_audiometric_2015}. To mitigate sounds containing such frequencies, we used a LPF with a cutoff frequency of 1kHz. This may help make certain trigger sounds more bearable while still preserving speech, which primarily falls below 1 kHz, excluding harmonics.

\begin{figure}[th]
    \centering
    \includegraphics[width=0.98\linewidth]{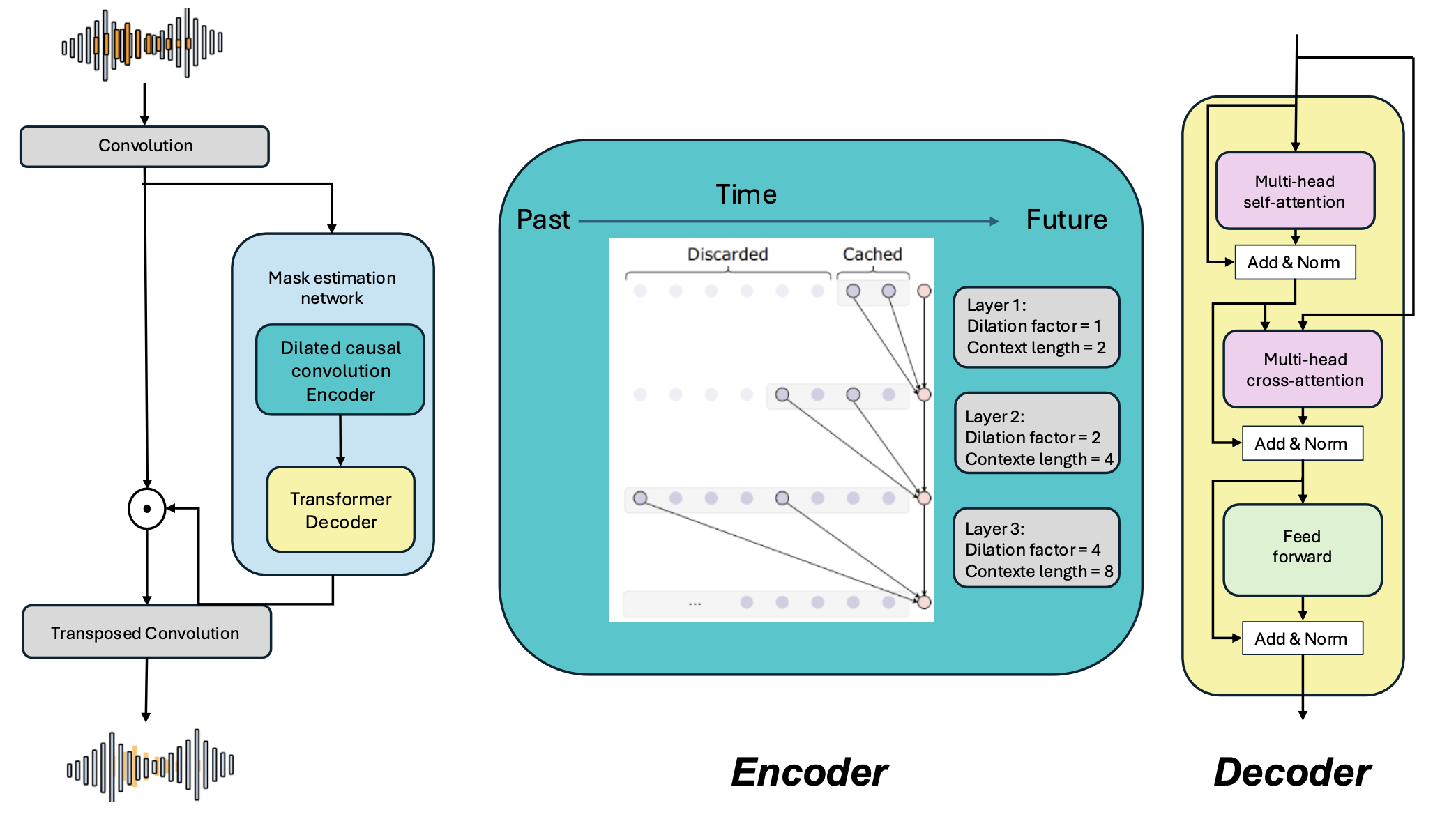}

   \caption{
      {ML-based Auto-encoder Architecture. The Encoder consists of 10 layers of dilated causal convolution. To increase the flexibility of the Semantic hearing model, we increased the latent space representation dimension to 512. The Decoder consists of self and cross-attention mechanisms, the former focusing on the temporal relationships within the decoder and the latter helping the model focus on relevant parts of the input audio, such as specific spectral or harmonic features that characterize the target sound.}}
    \label{fig:autoencoder}
    
\end{figure}

\subsubsection{ML-based auto-encoder algorithm}

In addition to DSP algorithms, we designed the auto-encoder model based on the Waveformer and Semantic Hearing models introduced by Veluri et al.~\cite{veluri_real-time_2023,veluri_semantic_2023}, which, despite its transformer-based architecture has promising real-time application due to fairly reasonable latency (6.56 ms). Unlike Veluri et al.\cite{veluri_semantic_2023}, no labels were provided as input to the Decoder. This decision was made to reduce model complexity and allow the model to automatically recognize trigger sounds. In addition, the model might learn to generalize triggering characteristics and identify the underlying features of a trigger.



The network, shown on Fig.\ref{fig:autoencoder}, was trained separately on \textbf{Dataset 1} and \textbf{Dataset 2}, and the resulting models will be referred to as NN1 and NN2, respectively. 
Similar to Veluri et al.~\cite{veluri_semantic_2023},  a negative SI-SNR \cite{roux_sdr_2019} was used as a loss function. One advantage of using SI-SNR is that it is invariant to the magnitude of the audio signal.
NN1 and NN2 were trained with 150 and 50 epochs, respectively, both with Adam optimizer and $5e^{-4}$ learning rate.

\section{Evaluation and results}
\label{sec:results}
To ensure that our \textbf{NN1} and \textbf{NN2} models are not biased on the respective test sets, we designed a new \textbf{Test Set 3} used for both objective and subjective assessment. The test contains 10 different 5-second stimuli, each with a different trigger and neutral sound pairing. The audio mixtures were constructed by combining three sounds over five seconds: a trigger sound (at least 0.5 seconds long) set at 0 dB, a neutral sound (at least 3 seconds long) set at -10 dB, and background traffic noise set at -35 dB.

\subsection{Objective evaluation}


The algorithm's performance was evaluated using difference metrics (\(\Delta\)-metrics) from SI-SNR \cite{roux_sdr_2019} to assess the improvement in an audio mixture after processing. \(\Delta\)SI-SNR represents the change in SI-SNR between the processed audio and the original mixture. A positive \(\Delta\)-value indicates enhancement, whereas a negative value suggests degradation, reflecting how much closer the processed audio is to the ground truth.

\begin{figure}[th]
    \centering

    \includegraphics[width=0.98\linewidth]{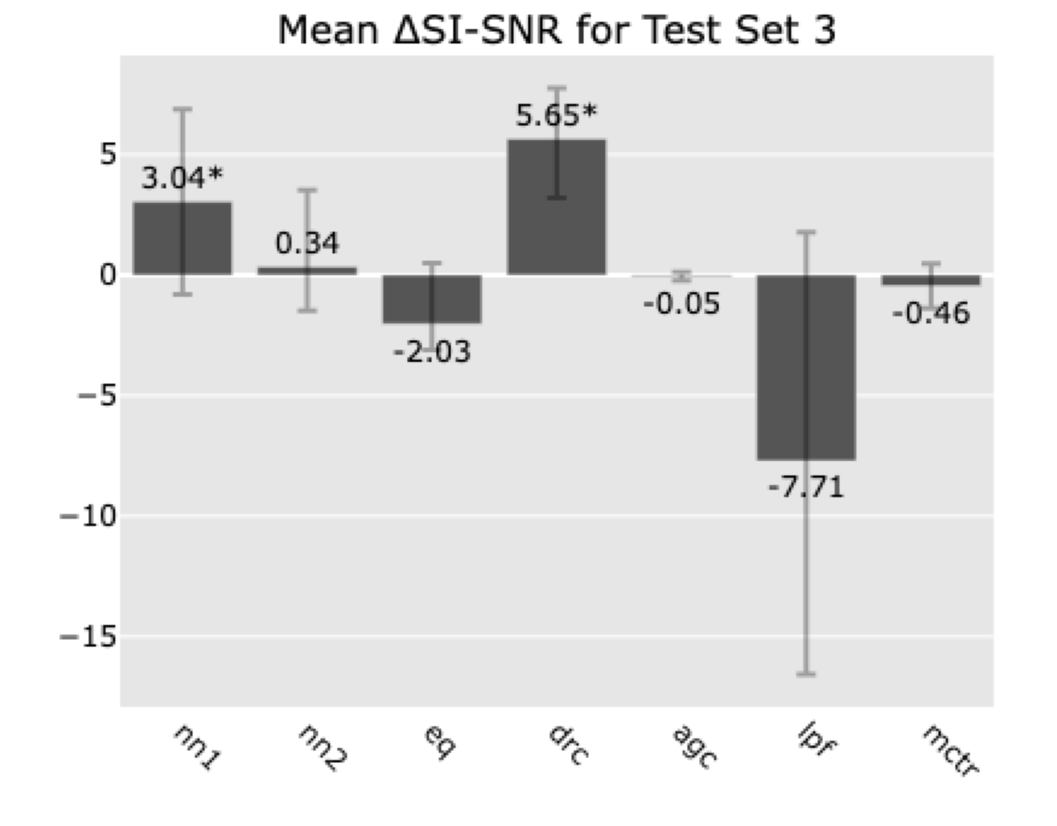}

   \caption{
      \centering
      {Objective performance in terms of SI-SNR \(\Delta\)-metric that represents the enhancement (positive values) and degradation (negative values).
      }}
    \label{fig:Seperation performance}
\end{figure}

Fig.~\ref{fig:Seperation performance} shows the objective performance results.
Both DRC and NN were able to attenuate transient trigger sounds to some extent, having positive $\Delta$ SI-SNR values and being the best-performing algorithms. However, both DRC and NN struggle with longer-lasting sounds and frequently do not cleanly separate trigger sources. Instead, when attenuating trigger sounds, they also often affect the rest of the mixture to some degree, which becomes evident when listening to a few samples on the project's webpage\footnote{\url{https://assistiveaudio.github.io/neurodivergent_audio/}}. 
The other algorithms either degrade (LPF), silence (AGC) or leave the audio mostly the same when not accounting for the volume (MCTR and AGC).

DRC appears to outperform NN significantly, particularly in the listening test (see next section), but also in the objective metrics. The much more pronounced difference between NN and DRC for the listening test could also be attributed to participants' sensitivity to audible distortions introduced by the NN. 

\subsection{Subjective evaluation}
\label{sec:subjective}

A listening test was conducted to evaluate how well various algorithms reduced triggerability in audio mixtures containing trigger sounds. The test simulated a possible application of audio enhancement algorithms as selective transparency mode in noise-cancelling headphones. Audio mixtures were first processed through a simulated noise-cancellation pipeline, using an attenuation curve of the SONY WH-1000XM5 noise cancellation headphones, including passive as well as active noise cancellation, and then added together with outputs from different algorithms, including DRC, NN1, and EQ, which had demonstrated strong objective performance.

\begin{table*}[ht]
    \centering
    \caption{Triggerability ratings ($\downarrow$) of N/C (Neurodivergent/Control) listeners. The best selective transparency system is shown in bold. Values marked with an asterisk indicate 0.001 significantly higher triggerability compared to control (see Fig.~\ref{fig:Seperation performance}). The superscript letters of the overall mean values indicate 0.05 significantly higher triggerability compared to the competing algorithms. For example, the \textbf{anc-eq} achieves significantly different results from \textbf{anc-nn} (n) and \textbf{anc-drc} (d) systems.}
    \label{tab:1}
    \resizebox{\textwidth}{!}{%
    \begin{tabular}{l|cc|cccccc}
    \toprule
    &  \multicolumn{2}{c|}{\textbf{mix}} & \multicolumn{2}{c}{\textbf{anc-eq}} & \multicolumn{2}{c}{\textbf{anc-nn}} & \multicolumn{2}{c}{\textbf{anc-drc}} \\
    \textbf{Trigger sounds} & N & C & N & C & N & C & N & C\\
    \midrule
    alarm  & 80.9 & 80.1 & 77.1 & 74.8 & 66.6 & 61.0 & \textbf{46.4} & \textbf{43.4}\\
    barking  & 54.4 & 55.9 & 52.6 & 55.6 & 35.7 & 31.5 & \textbf{30.7} & \textbf{28.7} \\
    breathing  & 68.6 & 68.4 & 68.6 & 69.8 & 49.3 & 43.6 & \textbf{39.9} & \textbf{33.4} \\
    chewing, -mastication  & 69.1 & 54.9 & 65.1 & 52.5 & 56.4 & 42.1 & \textbf{51.1} & \textbf{37.9} \\
    cutlery, -silverware  & 67.9 & 56.1 & 66.2 & 54.6 & 40.9 & 33.4 & \textbf{35.8} & \textbf{29.0} \\
    finger-snapping & 55.9 & 52.1 & 53.7 & 54.9 & \textbf{37.3} & 32.5 & 37.5 & \textbf{32.2}\\
    slamming & 61.9 & 61.7 & 61.5 & 62.0 & 37.1 & 32.9 & \textbf{23.5} & \textbf{25.1} \\
    sniffing & 70.2 & 71.3 & 69.8 & 72.6 & 43.7 & 39.0 & \textbf{39.8} & \textbf{37.4} \\
    squeaking & 74.8 & 76.7 & 76.6 & 75.6 & 63.3 & 60.6 & \textbf{39.4} & \textbf{33.9} \\
    tapping & 58.0 & 53.1 & 53.9 & 48.4 & \textbf{36.2} & 24.9 & 36.9 & \textbf{29.7} \\
    \midrule
    \textbf{Overall mean} & $^{*}$66.71$^{e, n, d}$ &63.09$^{e, n, d}$& 64.67$^{n, d}$ &62.14$^{n, d}$& $^{*}$46.86$^{d}$ &40.15$^{d}$& $^{*}$38.22$^{}$ &33.13$^{}$\\
    \bottomrule
    \end{tabular}
    }
\end{table*}

Participants completed the test using the SenseLabOnline platform\footnote{\url{https://senselabonline.com}}, rating the 10 different 5-second mixture, each with a different trigger and neutral sound pairings. For a given mixture, each processed stimulus, alongside an unprocessed (original) mixture and a ground truth version where the trigger was completely removed, was displayed at the same time on the same page and rated along with each other. These processed versions were anonymized, and their presentation order was fully randomized for each participant to eliminate bias and order effects. 
The listening was conducted by neurodivergent individuals as well as a control group, which consisted solely of neurotypical participants. Participants rated the extent to which each processed mixture elicited a negative response (triggerability) using a continuous scale from 0 associated with ``very weak'' to 100 associated with ``very strong.'' In total, 133 neurodivergent and 47 control participants took part in our listening test. Tab~\ref{tab:1} shows the triggerability ratings of both neurodivergent and control group listeners. 

The listening test confirms that neurodivergent individuals have heightened sensitivity to auditory triggers compared to the general population, consistently reporting higher trigger scores across all processing methods.
While many sounds were triggering for both groups, some were specifically more distressing for the neurodivergent individuals. In particular, ``chewing, mastication'' and ``cutlery, silverware'' were rated significantly higher by the neurodivergent participants, with their scores well above 50, while the control group’s ratings hovered closer to this threshold.
However, many other trigger sounds were also distressing to the general population, indicating that certain sounds are generally uncomfortable, not just for neurodivergent individuals. Still, the significantly higher ratings for some trigger sounds in the neurodivergent group suggest that these individuals experience a more intense and distinct reaction to particular sounds.

\begin{figure}[th]
    \centering

    \includegraphics[width=1\linewidth]{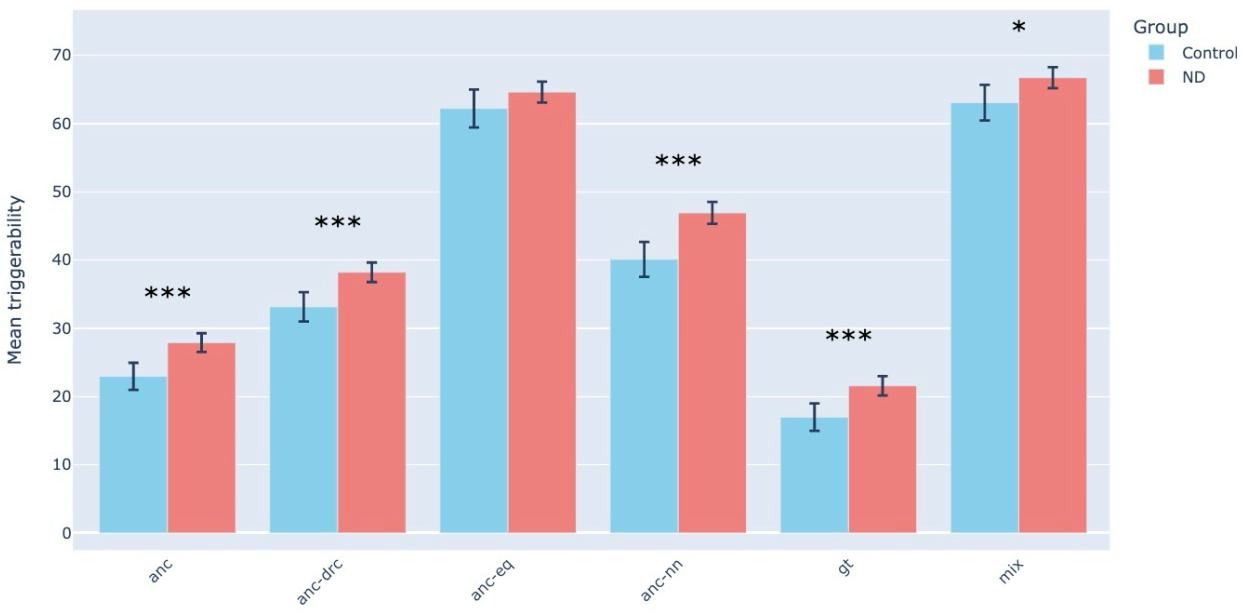}

   \caption{
      {Triggerability Comparison with Control. T-bars represent bootstrapped 95\% confidence intervals. Stars indicate the significance level of a group having a higher triggerability compared to the counterpart ($\text{*}p < 0.05$, $\text{**}p < 0.01$, $\text{***}p < 0.001$). For the comparisons, t-tests were conducted and the $p$-values were adjusted using Benjamini-Hochberg correction.
      }}
    \label{fig:Seperation performance}
    
\end{figure}

\section{Conclusion and future work}
\label{sec:conclusion}

This study has shown the potential of low-latency assistive audio enhancement in reducing auditory distress for neurodivergent individuals by selectively attenuating trigger sounds and its potential use in a selective transparency mode. A key aspect of this research was the creation of a trigger sound dataset, which enabled the training and evaluation of both DSP and ML audio enhancement algorithms. The data will be open after the conference decision. Among these algorithms, DRC that has low algorithmic latency emerged as the most effective in attenuating trigger sounds and reducing triggerability. The second best method was low-latency semantic hearing model.


The performance of algorithms like DRC depends on the SNRs of individual sources in the mixture and whether the trigger is in the background or foreground. Thus, conducting listening tests at varying SNR levels would be beneficial. Listening tests should be conducted to determine to what extent neutral or non-triggering sounds remain recognizable after being processed by audio processing algorithms such as DRC or neural networks.
Since triggers often appear in the foreground -- otherwise, they would hypothetically blend into the background and be less triggering -- DRC could be a viable low-latency solution for a selective transparency mode that filters out particularly loud or transient triggers. It could also be paired with a neural network that dynamically adjusts its parameters, further enhancing its capabilities.

Concerning the ML approach, which seems to be promising but introduces distortions, 
overlapping sounds frequently pose challenges in target sound separation, with performance often hinging on the size of the training dataset and the chosen network architecture~\cite{nath_separation_2024}. Therefore, increasing the amount of training data and employing a more flexible network may lead to better generalization and cleaner separation. Indeed, relying on only five hours of trigger sounds and approximately 55 hours of total training data might have been insufficient. Incorporating additional data augmentation strategies, such as time-shifting and pitch-shifting, could further improve results. Although hyperparameter tuning is also an option, it may be computationally expensive and thus requires careful consideration.
Moreover, the models are trained on data containing only a single trigger type per audio sample. However, it might be valuable to develop algorithms that can handle multiple trigger sounds simultaneously. In contrast, DRC may have the capability to effectively process multiple transient triggers in a single audio sample.



\section{Acknowledgments}

We would like to express our heartfelt gratitude to the pool of neurodivergent listening test participants for their time and effort.

\vfill\pagebreak


\bibliographystyle{IEEEbib}
\bibliography{referencesmc,mybib}

\end{document}